\documentclass[12pt,a4paper]{article}
\usepackage[latin1]{inputenc}
\usepackage{amsmath}
\usepackage{amsfonts}
\usepackage{amssymb}
\usepackage{graphicx}
\usepackage{wrapfig}
\usepackage[a4paper,margin=1.0in]{geometry}
\usepackage{hyperref}

\begin{document}

\hyphenation{emit-tance de-po-si-tion}

\title{Undulator-Based Positron Source at 250 GeV CM Energy with Different 
	Optical Matching Devices: Pulsed Flux Concentrator and Quarter Wave
	Transformer\thanks{Talk presented at the International Workshop on Future Linear
	Colliders (LCWS2017), Strasbourg, France, 23-27 October 2017. C17-10-23.2.}}

\author{Andriy Ushakov$^1$\thanks{ andriy.ushakov@uni-hamburg.de},
	Gudrid Moortgat-Pick$^{1,2}$ and Sabine Riemann$^3$\\
 \\ $~$
 \small $^1$\textit{University of Hamburg, Luruper Chaussee 149, D-22761 Hamburg, Germany} \\
 \small $^2$\textit{Deutsches Elektronen-Synchrotron (DESY), Notkestrasse 85, D-22607 Hamburg, Germany}\\
 \small $^3$\textit{Deutsches Elektronen-Synchrotron (DESY), Platanenallee 6, D-15738 Zeuthen, Germany}}

\date{}

\maketitle{}

\begin{abstract}
In the baseline design of the International Linear Collider (ILC) an
undulator-based source is foreseen for the positron source. In this study the
energy deposition in the pulsed flux concentrator (FC) of positron source is
calculated for the ILC at 250 GeV center-of-mass energy. The peak energy of 
33~J/g deposited by one beam pulse in the current design of the FC is above the
limit for copper material. Several promising options were considered to solve
the issue of overheating the FC: the quarter wave transformer (QWT) has a
significantly bigger aperture and is considered as an valuable and safe
alternative for the FC. Since the positron source with a QWT is expected to lead
to a lower positron capture efficiency, also the expected positron yield was
calculated in addition to the energy deposition in QWT.
\end{abstract}

\section{Introduction}

The baseline positron source for the future International Linear Collider (ILC)
is based on a superconducting helical undulator that is placed at the end of
main linear accelerator~\cite{ref:TDR} and produce a large amount of
circularly-polarized photons hitting the target. The efficiency of positron
generation in a thin titanium alloy by undulator photons depends strongly on the
energy of electron drive beam passing the undulator. The average photon energy
is 10.7 MeV for a 150 GeV drive beam and an undulator with 11.5 mm period and
0.86~T field on beam axis (i.e. $K = 0.92$). The space reserved for the
undulator is 231 meters. In order to get 1.5 positrons per initial electron at
the damping ring (DR) for an undulator with an 150 GeV drive beam, requires an
undulator magnet length of about 147~meters. One achieves a photon power of
64~kW for $2\cdot10^{10}$ e$^-$/bunch, 1312~bunches/pulse and 5 Hz pulse
repetition rate.

For the center-of-mass (CM) energy of 250~GeV, i.e. for  an 125 GeV electron
drive beam but the same undulator parameters as for 150 GeV case, the average
photon energy is 7.4 MeV. In order to get the same yield as before, 1.5
e$^+$/e$^-$, all 231 meters have to be used for undulator magnets. The average
photon power is increased by about 10\% (70.5 kW). The calculations of the
positron yield have shown that the required positron current can also be
achieved at 250~GeV CM energy by optimizing the positron source
parameters~\cite{ref:IPAC13,ref:AWLC17}. The lower energy of photons, the higher
photon power and the bigger opening angle of the undulator synchrotron radiation
result in the higher energy deposition in an optical matching device (OMD)
downstream the target and in the accelerator structures. Therefore, at low
energies, the possible issue of overheating the OMD and the accelerator
structures of the capture section has to be checked carefully. The energy
deposition in two different types of OMD (the pulsed FC and QWT) is considered
in this paper as well.

The Geant4-based application code, PPS-Sim \cite{ref:pps-sim}, was used for the 
calculations of the positron yield and for evaluating which undulator length and/or
field are required to achieve the positron yield of 1.5~e$^+$/e$^-$.
In the yield estimations, the ILC DR longitudinal acceptance
($\pm 3.4\,\mathrm{cm} \times \pm 37.5\,\mathrm{MeV}$) and the transverse
acceptance (sum of normalized $x$- and $y$- emittances of 70 mm$\cdot$rad) were
taken in account. The calculations of energy deposition and radiation damage in
FC and QWT were performed in FLUKA \cite{ref:fluka}.

\section{Pulsed Flux Concentrator}

The optical matching device (FC or QWT) is used to match the positron beam
emerging from the target and characterized by small beam size and large divergence
into the beam with small angular divergence and large size at the beginning
capture accelerator.
The FC is a pulsed tapered magnet. It was designed and prototyped at Lawrence Livermore
National Laboratory.
The FC has demonstrated a very good stability of fields in space and time~\cite{ref:Gronberg12}.
The maximal field of the FC is 3.2 T at 2 cm from the rear side of target was used
in simulations. The smallest aperture radius of the FC is 6.5 mm. The distance
between Ti6Al4V target having 0.4 radiation length thickness and the 1st copper
concentrating plate of the FC is 8 mm. 

\begin{figure}[htb]
  \centering
  \includegraphics*[width=80mm]{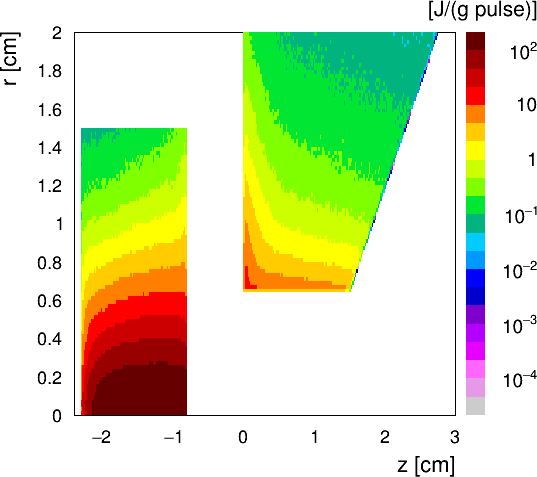}$~$
  \includegraphics*[width=76mm]{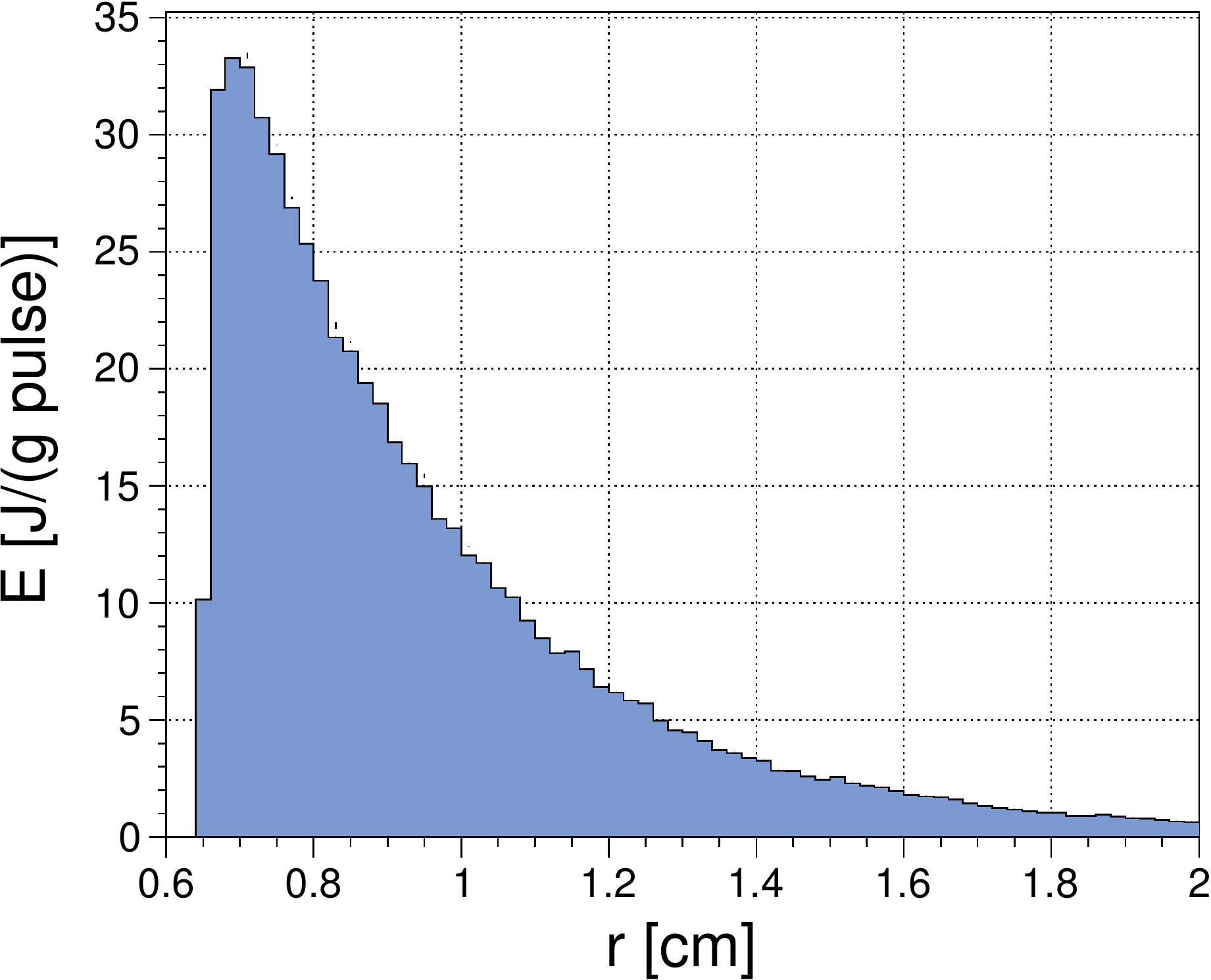}
  \caption{The energy deposition in the target and at the 3.2 T pulsed FC with 6.5 mm aperture radius (left). The radial profile of the deposited energy near the front surface of the FC (right).}
  \label{fEdepInFC}
  \end{figure}

The distribution of the energy deposited by the beam in the stationary target
and in the part of FC near the target is shown in the left plot of
Figure\,\ref{fEdepInFC} for the positron source with the following parameters:
125 GeV electron drive beam with 1312 bunches/pulse, 231~m undulator length with
$K = 0.85$ and 570 m distance between the middle of the undulator and the
target. The issue of very high peak energy deposition density (PEDD) in the
target is reduced to a tolerable level by the proper choice of the size and the
rotation speed of the target. Taking into account that the beam heat at the
target (with one meter diameter) near the outer edge and with the tangential
speed of 100 m/s, the PEDD in the target is 44 J/(g pulse). The upper limits of
PEDD and temperatures for the target and the OMD materials should be checked
experimentally under conditions that are close to those expected at the ILC.
Such experiments have already been started and first results have already been
published~\cite{ref:IPAC17,ref:AWLC17SR}.

\begin{figure}[htb]
  \centering
  \includegraphics*[width=85mm]{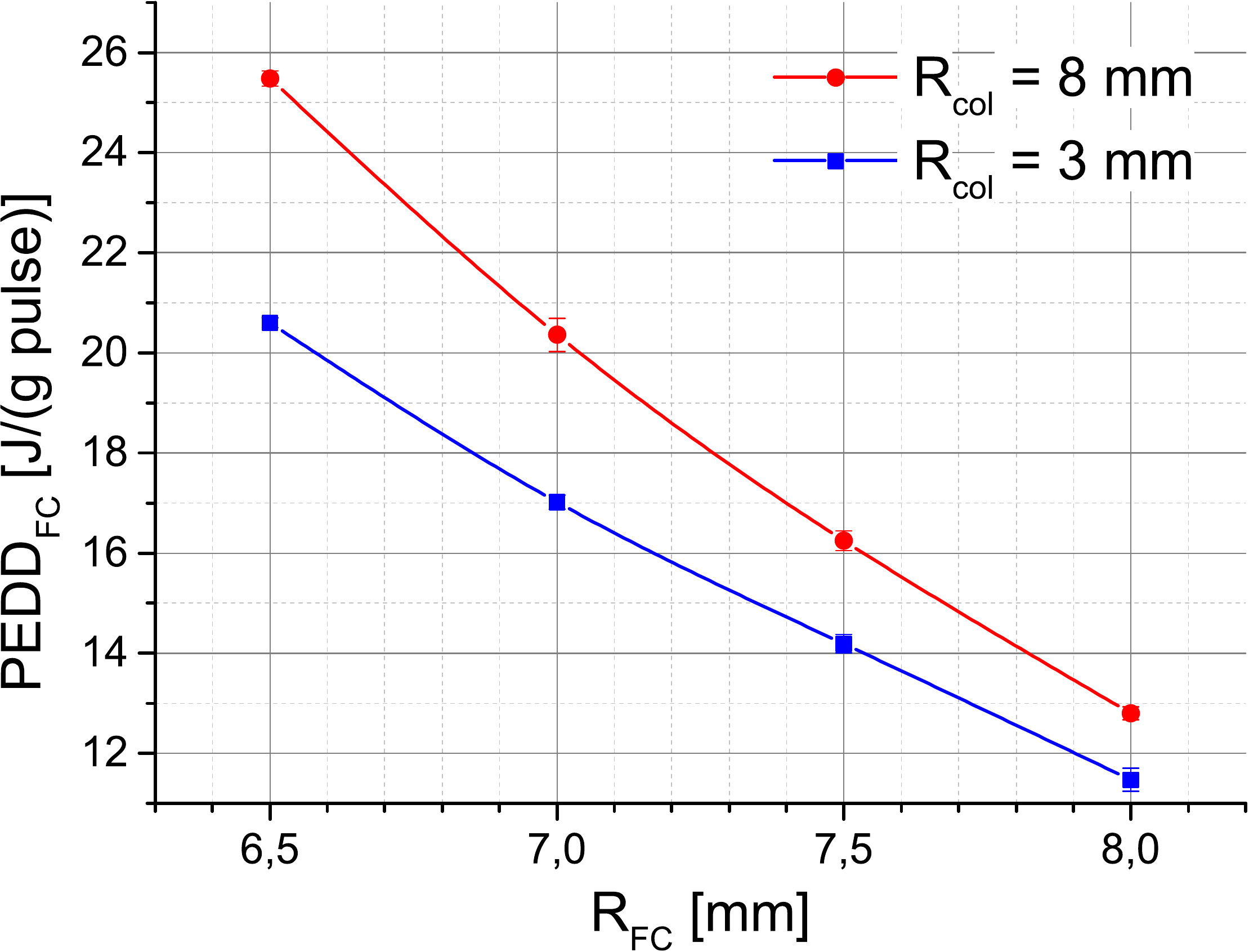}
  \caption{Peak energy deposition density in the FC with different aperture radii for the case of using photon collimators with 3 mm and 8 mm aperture radii, respectively.}
  \label{fPEDDfc-vs-Rfc}
\end{figure}

The radial profile of the energy deposited in hottest area of the FC (i.e. near
the front surface) is shown in Figure~\ref{fEdepInFC} (right). The estimated
$\mathrm{PEDD_{FC}}$ was 33.3 J/(g pulse). This value is above the limit for
copper ($7\div 12$~J/g \cite{ref:XFEL}). In order to reduce
$\mathrm{PEDD_{FC}}$, several promising modifications of positron source
parameters have been simulated \cite{ref:AWLC17} and are shortly summarized
below:
\begin{enumerate}
\item Using a compact electron dogleg only for the 125 GeV electron drive beam
instead of the TDR dogleg, allows to shorten the distance between the middle of
undulator and the target by 168.8 meters (from 569.9~m down to 401.1~m)
\cite{ref:OkugiWG2}. That makes the spot size of photons on the target smaller
and results in a reduction of $\mathrm{PEDD_{FC}}$ from 33.3 J/(g pulse) down to
29.3 J/(g pulse) for a 231 m-long helical undulator with $K = 0.85$ and an 125
GeV e$^-$ drive beam.
\item The target thickness can be reduced to 7 mm without any losses in the
positron yield \cite{ref:AWLC17}. It makes target cooling easier and results in
approximately 8\% lower $\mathrm{PEDD_{FC}}$.
\item Taking into account the 3 GeV energy loss of the electrons along the
undulator (128~GeV e$^-$ beam at beginning of undulator) and the change of the
target thickness down to 7 mm results in $\mathrm{PEDD_{FC}} = 25.5$ J/(g
pulse).
\item In addition to the already mentioned changes, the application of a photon
collimator with an aperture radius of 2.5 mm upstream target can reduce the PEDD
down to 18.5~J/(g pulse).
\item In order to reduce $\mathrm{PEDD_{FC}}$ down to 12 J/(g pulse), the
aperture radius at the beginning of the FC ($R_{\mathrm{FC}}$) has to be
increased from 6.5 mm to approx. 8 mm for the case of the changes mentioned
above plus a 3 mm aperture radius of the photon collimator. The dependence of
$\mathrm{PEDD_{FC}}$ on $R_{\mathrm{FC}}$ is shown in Figure
\ref{fPEDDfc-vs-Rfc}. The Figure includes, in addition to the data for a photon
collimator with  $R_{\mathrm{FC}} = 3$ mm, also the curve for $R_{\mathrm{FC}} =
8$~mm. The photon collimator with $R_{\mathrm{FC}} = 8$ mm has no impact on the
positron yield or positron current at the DR \cite{ref:AWLC17}, i.e. concerning
the yield, such a large aperture collimator is equal to the case without
collimator.
\end{enumerate}         

\begin{figure}[htb]
  \centering
  \includegraphics*[width=88mm]{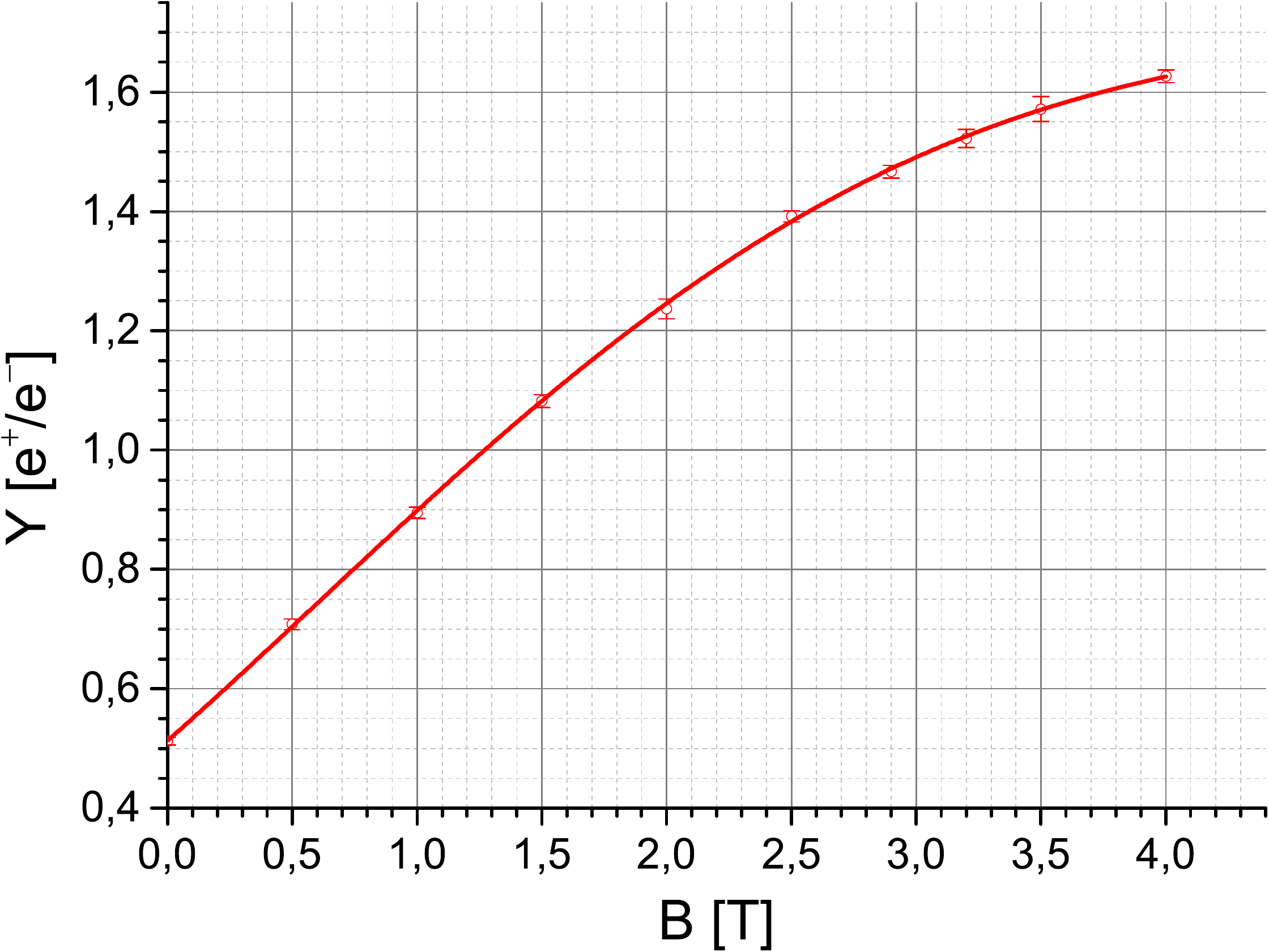}
  \caption{Dependence of positron yield on maximal field of FC.}
  \label{fYfc-vs-Bmax}
\end{figure}

In our calculations, it was supposed, that all FC with different apertures have
a 3.2~T magnetic peak field on the beam axis ($B_{\mathrm{FC}}$). If the peak
field is challenging for a FC with an 8 mm aperture radius (or larger) the
$B_{\mathrm{FC}}$ will be lower and the positron yield as well. The typical
dependence of the yield on $B_{\mathrm{FC}}$ at 250 GeV CM energy is shown in
Figure \ref{fYfc-vs-Bmax}. In the case that the reduction of the yield will be
unacceptable low, an alternative OMD has to be used. The positron yield in the
source using QWT and the energy deposition in QWT is considered below.

\section{Quarter Wave Transformer}

The QWT consists of two solenoids: the first one has a short length and a high
magnetic field, the second one has a long length, low magnetic field
superimposed on the accelerating electric field \cite{ref:Chehab}. The QWT for
the ILC positron source was considered by the group from Argonne National
Laboratory (ANL).
The estimated capture efficiency of positrons in the source using a QWT with
approximately 1 Tesla first QWT magnet downstream the target was only 25\% less
than in the source with a pulsed FC \cite{ref:WanmingQWT}. Both OMDs were using
the same 0.5 T second solenoid in which the accelerator structures were
embedded. The lower field of the QWT reduces the eddy currents in the rotated
target. DC QWT is easier to engineer in comparison to the FC that has to keep
constant (in time and space) during 1 ms pulse the eddy currents in the volume
close to the tapered aperture of FC.

\begin{figure}[htb]
	\centering
	\includegraphics*[width=65mm]{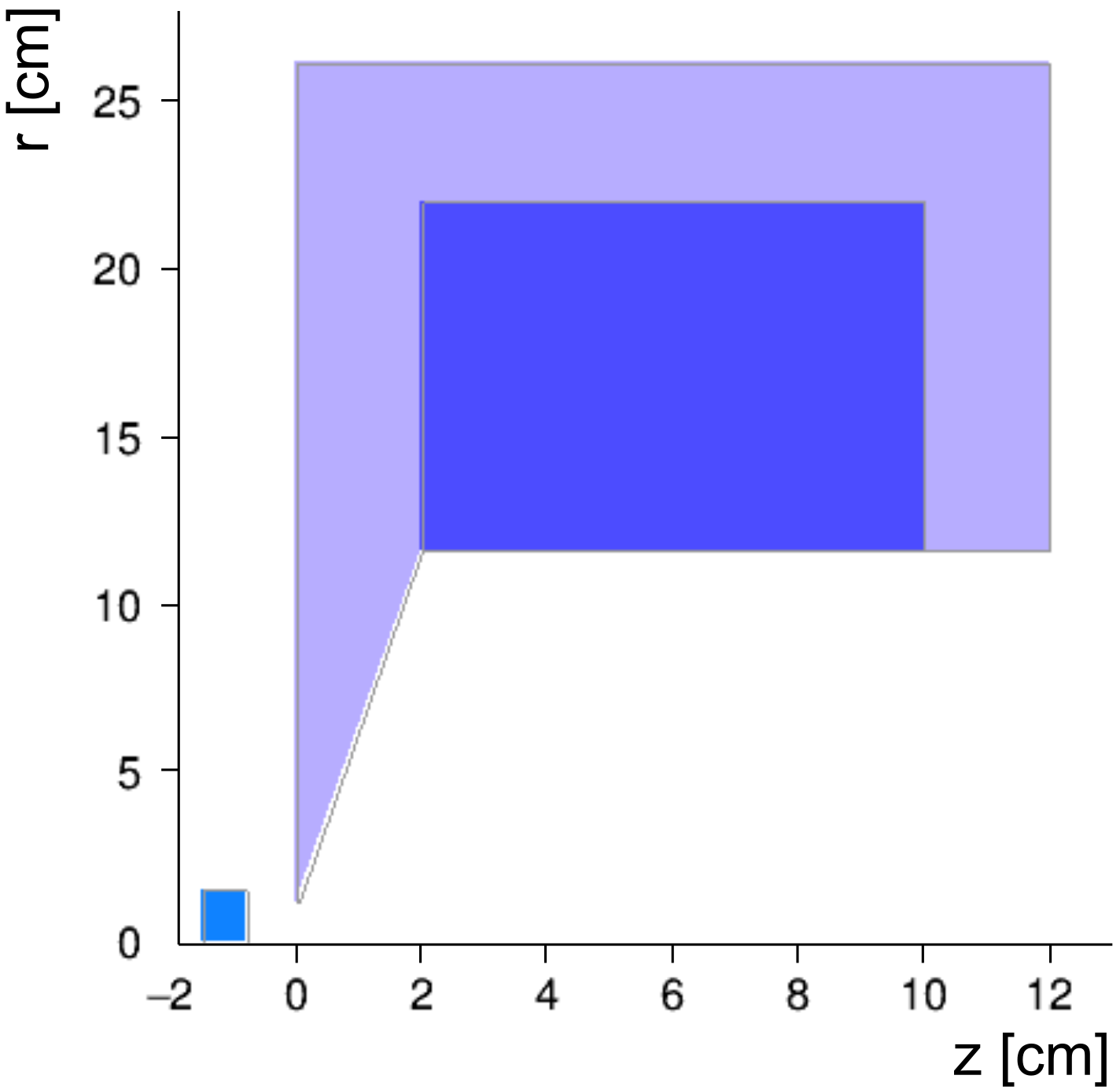}$~~~$
	\caption{Dimensions of 1st QWT magnet and position related to target.}
	\label{fQWTgeometry}
\end{figure}

The design for the QWT considered by the ANL group was used also in our
simulations. The first QWT magnet has 1.04 T field on the beam axis. It was
placed 8 mm from rear side of the target. The Figure \ref{fQWTgeometry} shows
the geometry and position of the first QWT magnet used in calculations. The coil
has 8 cm length and inner radius of 11.6 cm. The tapered iron core on target
side has changing radius from 1.1 cm to 11.6 cm in 2 cm in beam direction. The
free space between 1.04 T and 0.5 T solenoids was 4 cm.

\begin{figure}[htb]
	\centering
	\includegraphics*[width=88mm]{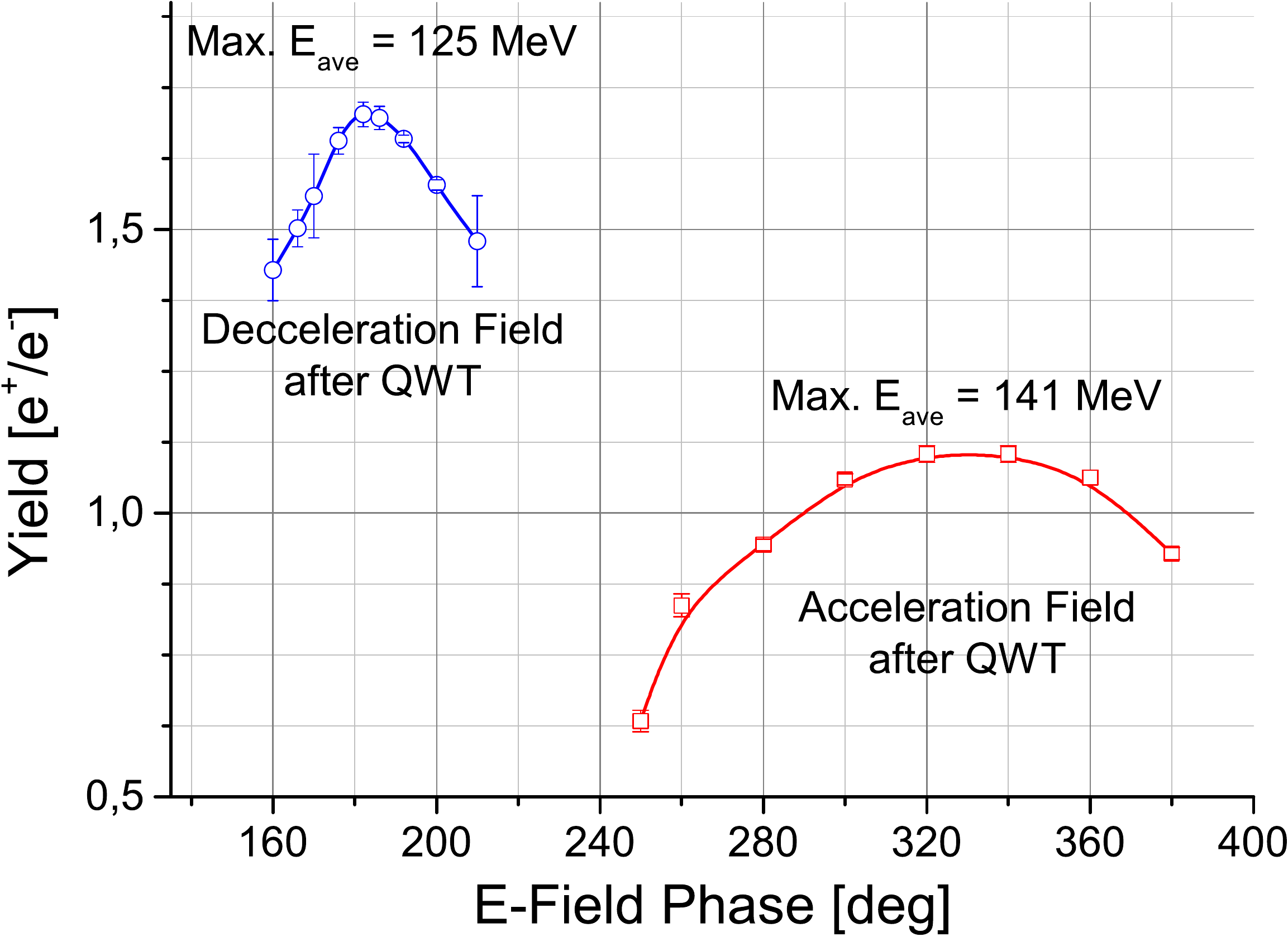}
	\caption{Dependence of positron yield on the phase of electric field at the beginning
		of capture accelerator.}
	\label{fYfc-vs-Ephase}
\end{figure}

The proper choice of the electric field phase at the beginning of capture
section has a big impact on the captured positron yield $Y_{e^+}$. The
dependence of $Y_{e^+}$ on the $E$-field phase is shown in Figure
\ref{fYfc-vs-Ephase} for the case of 250 GeV CM energy, TDR electron dogleg, 231
m undulator length with $K = 0.92$ and 7 mm target thickness. The deceleration
of positrons after the QWT helps to increase $Y_{e^+}$ significantly.

\begin{figure}[htb]
  \begin{center}
  \includegraphics*[width=78mm]{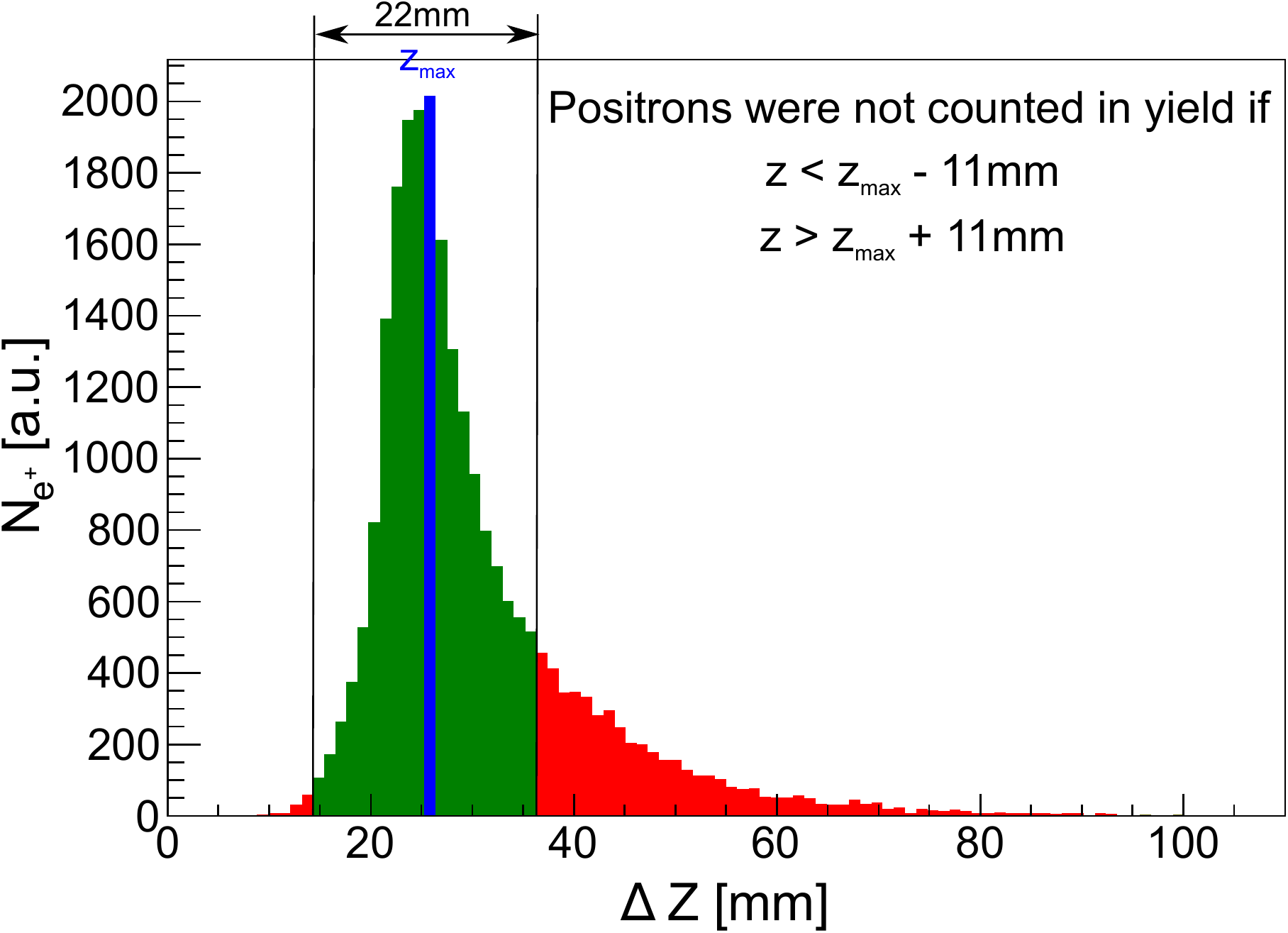}$~$
  \includegraphics*[width=78mm]{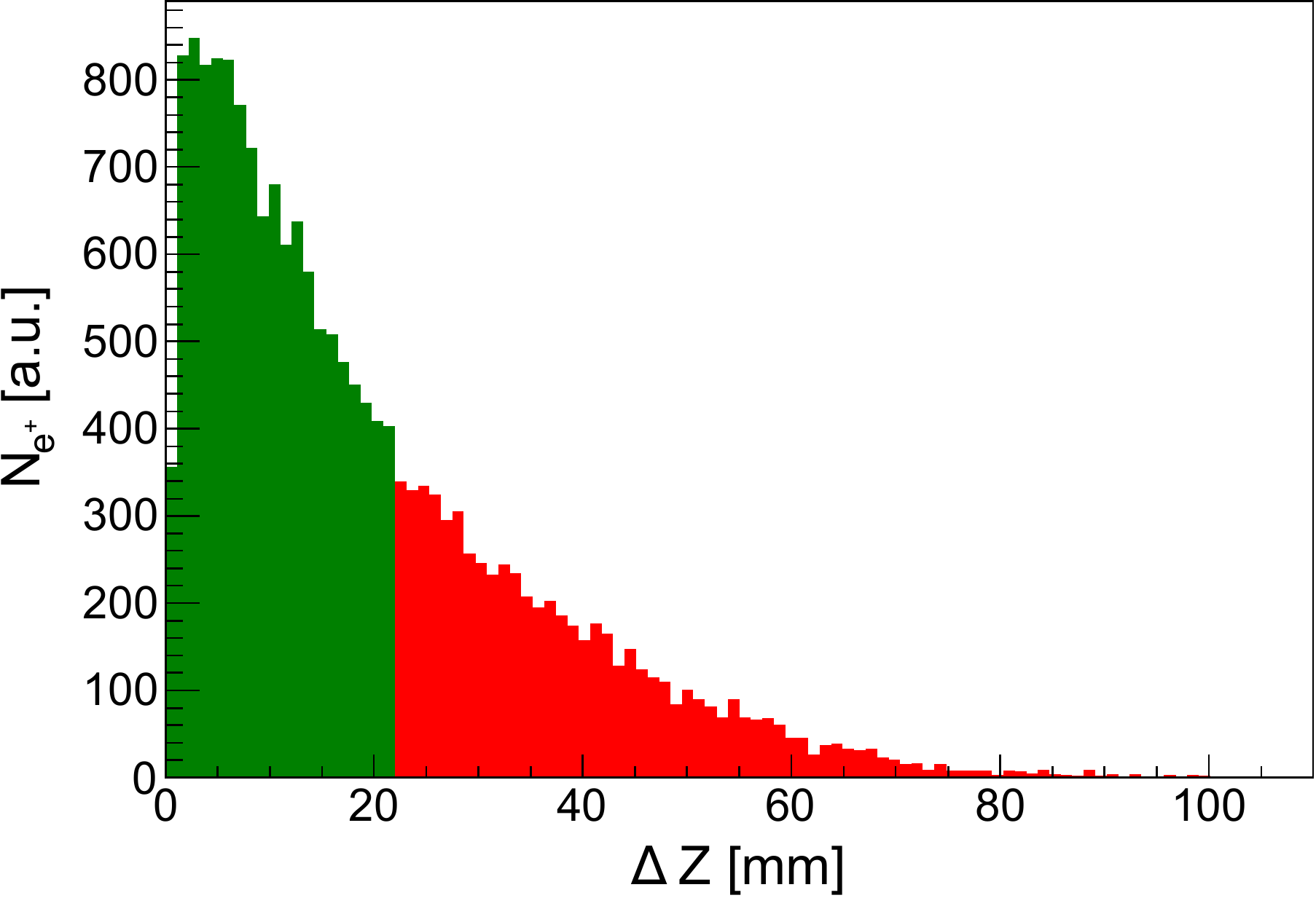}
  \caption{Longitudinal profiles of the positron density in the bunches in case of decelerating electric field at the beginning of the capture linac (left) and in case of an accelerating field (right). The fraction of positrons that satisfy the DR acceptance is shown in green, the positrons beyond DR acceptance are shown in red.}
  \label{fLongBunchProfiles}
  \end{center}
\end{figure}

The longitudinal profile of the positron density in the bunches at the end of
the capture section (125 MeV) in case of decelerating and accelerating electric
fields is shown in Figure \ref{fLongBunchProfiles}. It indicates that the
deceleration of positrons downstream the QWT results in a smaller bunch length.
The energy spread of positron beam is correlated with the bunch length. The ILC
energy acceptance of the damping ring (DR) is $\pm 0.75$\%. An energy compressor
system (ECS) is foreseen in ILC TDR and it is located downstream 5 GeV positron
booster in the positron-linac-to-ring part of source beam line together with a
spin rotator. According to the table with positron source parameters
\cite{ref:ParamTable}, the ECS is able to reduce the relative energy spread from
4.4\% to 1.5\%. The relation between the energy spread $\Delta E/E$ and the
bunch length $\Delta z_b$ can be described as $\Delta E/E = 1 - \cos(\omega
\Delta z_b / 2c)$, where $\omega$ is the RF frequency and $c$ is the speed of
light. An energy spread of $\pm 2.2$\% corresponds to $\pm 11$ mm of the bunch
length. The resulting fraction of positrons counted in the positron yield (green
color in Figure~\ref{fLongBunchProfiles}) has a full width of 22 mm.

\begin{figure}[htb]
	\centering
	\includegraphics*[width=84mm]{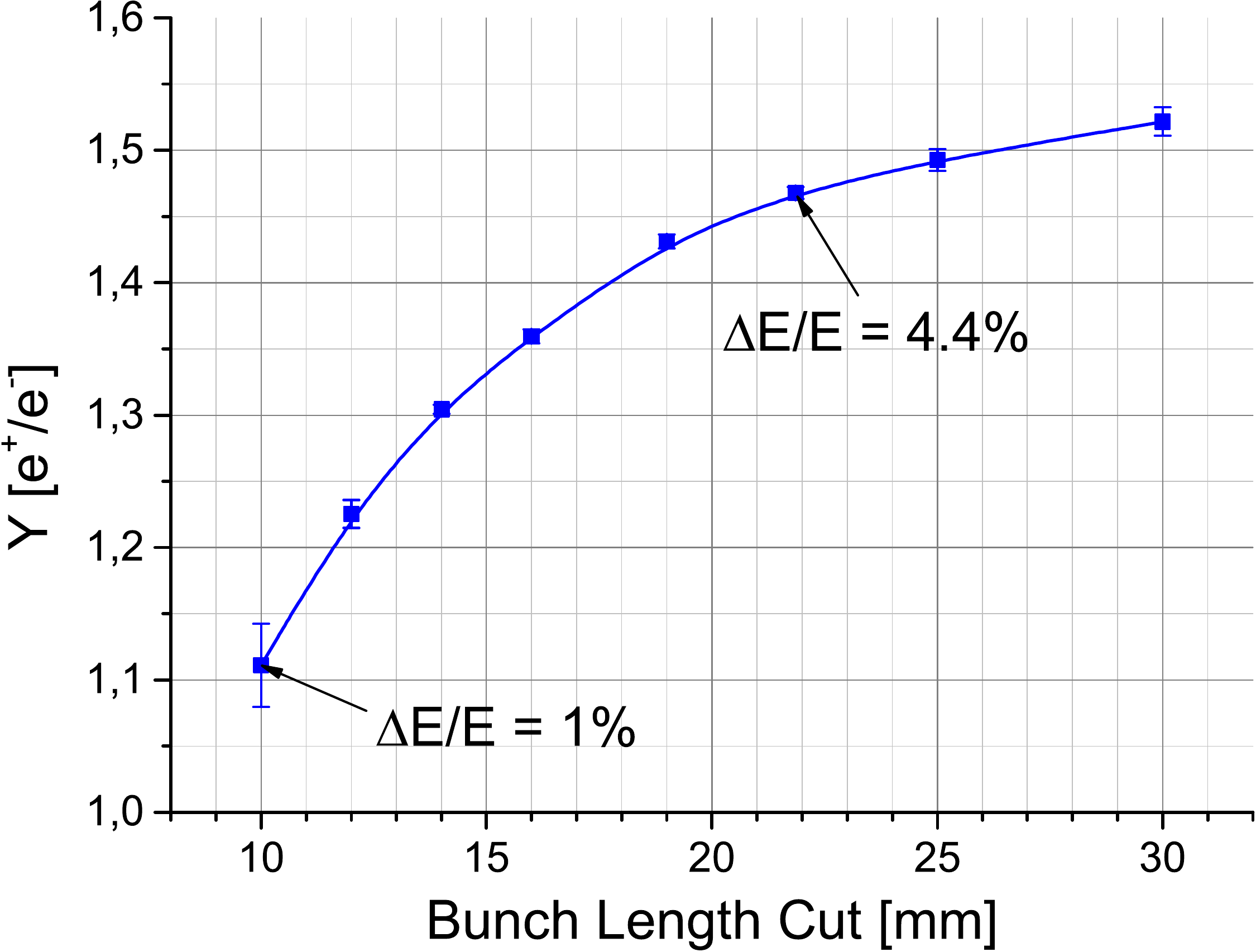}
	\caption{Impact of bunch length cuts at the end of the capture linac (125 MeV) on the positron yield. Relative energy spreads for 10 mm and 22 mm are shown by arrows.}
	\label{fYqwt-vs-dZbunch}
\end{figure}

The impact of different bunch length cuts applied at the end of capture linac
(125~MeV) on the positron yield is shown in Figure~\ref{fYqwt-vs-dZbunch}. The
relative energy spread $\Delta E/E$ of 1\% corresponds to $\Delta z_b = 10$~mm.
The 3.5 times energy compression was proposed by Y.~Batygin \cite{ref:Batygin}.
In our calculation of the capture yield, the more conservative compression from
4.4\% to 1.5\% was supposed and the bunch length cut of 22~mm was applied at the
end of capture section in addition to the 70 mm$\cdot$rad for the sum of
normalized $x$- and $y$- emittances.   
       
\begin{figure}[htb]
  \centering
  \includegraphics*[width=85mm]{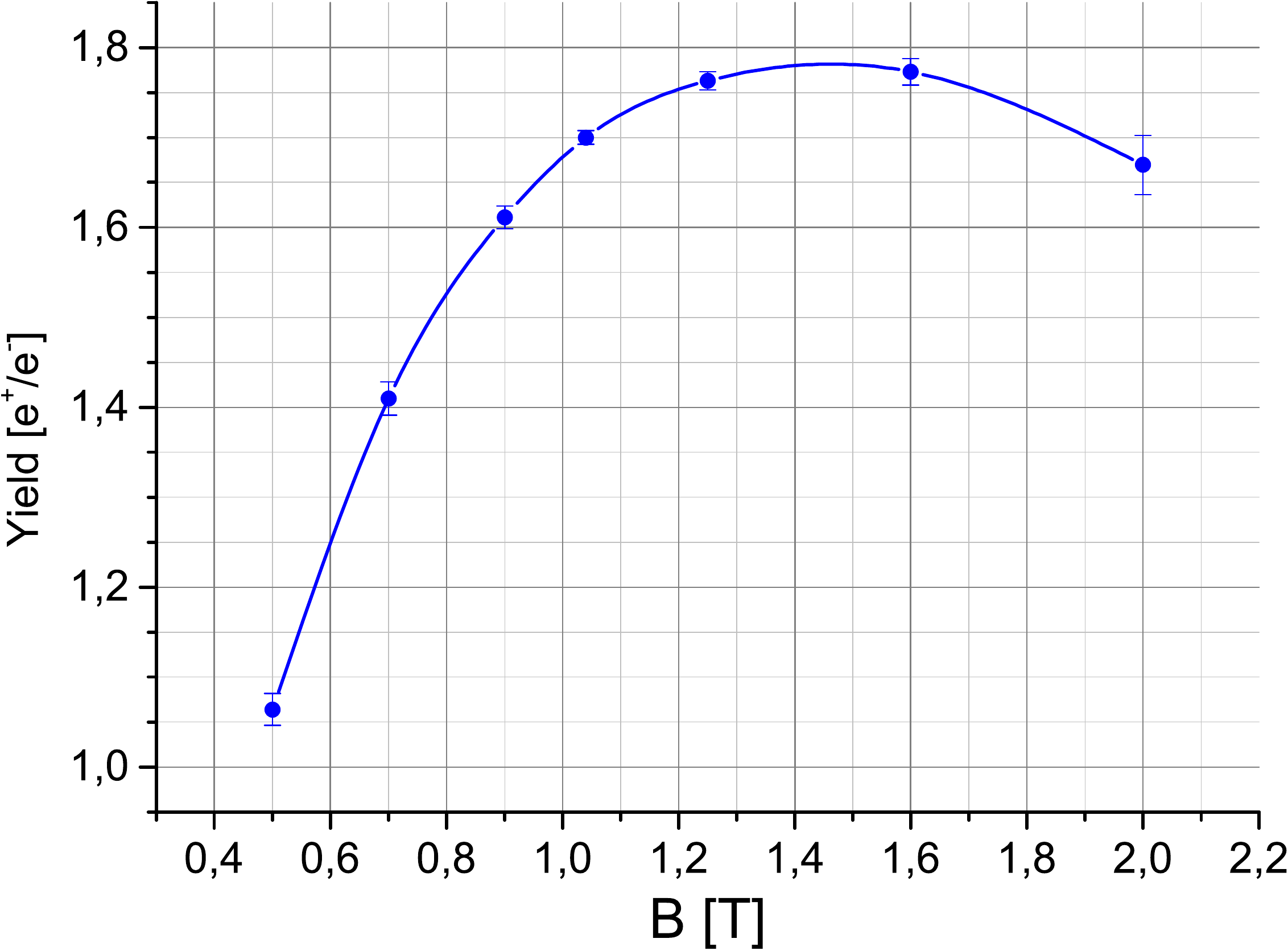}
  \caption{Dependence of positron yield on magnetic field strength of the 1st QWT magnet.}
  \label{fYqwt-vs-B}
\end{figure}

The dependence of the positron yield on the maximal field of the 1st QWT solenoid is
shown in Figure \ref{fYqwt-vs-B}. The optimal field for the 12 cm total length of 1st
solenoid is approximately 1.4~T. The positron yield for the suggested 1.04 T QWT
field is only a few percent less than at~1.4~T.

\begin{figure}[htb]
  \centering
  \includegraphics*[width=85mm]{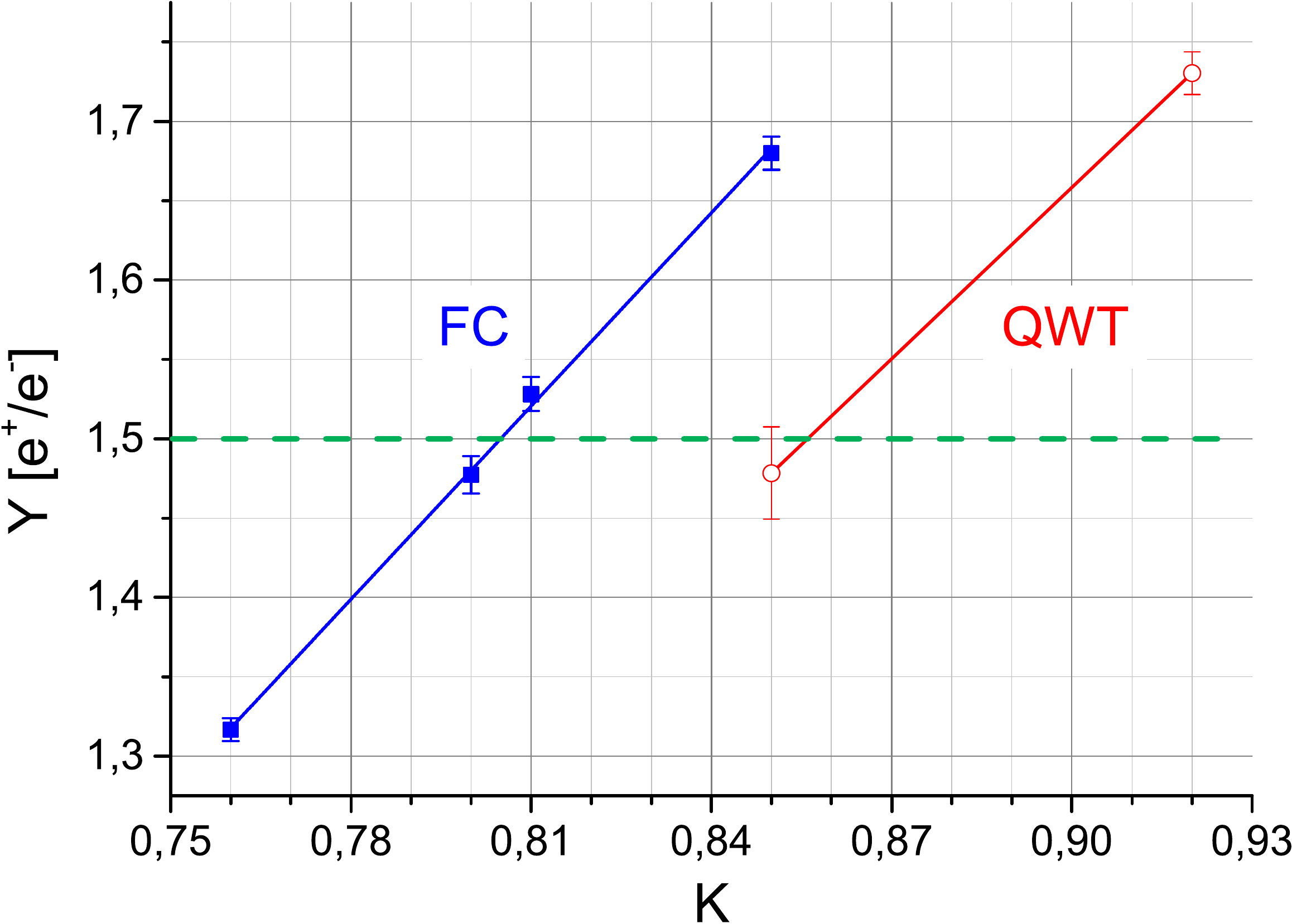}
  \caption{Positron yield of source with 3.2 T FC and 1.04 T QWT for 231 m
long helical undulator and different undulator $K$ values.}
  \label{fYfcYqwt-vs-K}
\end{figure}

The yield of the positron source with a 3.2 T FC and a 1.04 T QWT for a 231 m
long helical undulator but different undulator $K$ values is shown in Figure
\ref{fYfcYqwt-vs-K}. For undulators with higher $K$ values, higher number of
photons $N_{ph}$ are generated ($N_{ph} \sim K^2$). The efficiency of the
positron capture in case of using a FC is higher, but, the yield of
1.5~e$^+$/e$^-$ for the source at 250~GeV CM energy can be achieved with a QWT
too via a moderate increase of the undulator $K$ value from 0.8 to 0.85.

\begin{figure}[htb]
  \centering
  \includegraphics*[width=80mm]{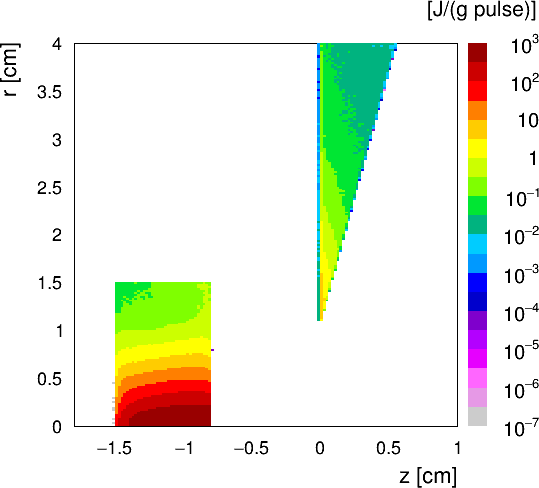}$~~~$
  \includegraphics*[width=71mm]{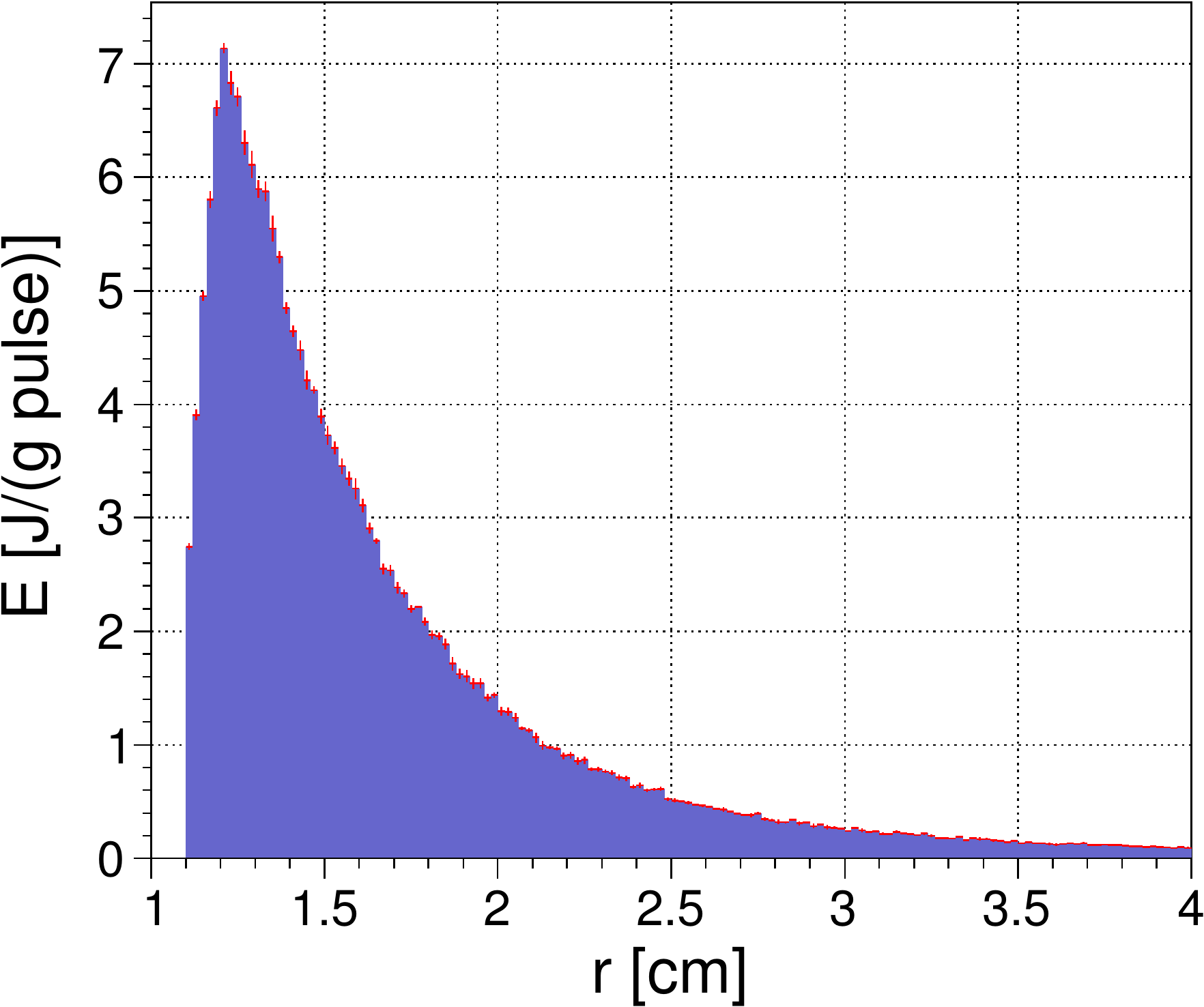}
  \caption{Distribution of the energy deposition (left) and the radial profile of energy deposition density close to the front surface of the 1st QWT magnet (right).}
  \label{fEdepQWT}
\end{figure}

The significantly bigger aperture size of the QWT in comparison to the FC (i.e.
11 mm in QWT vs 6.5 mm in FC) for approx. the same 8 mm distance between the
OMDs and the target leads to a large difference in the PEDD in the OMDs. The
energy distribution and PEDD in the QWT are shown in Figure \ref{fEdepQWT}. The
PEDD$_{\mathrm{QWT}}$ in an iron yoke of the 1st solenoid is 7~J/(g pulse). The
upper limit per pulse for the ARMCO Pure Steel is 12.5~J/g. The limit was
estimated on the base of a simplified approach used in TESLA FEL report
\cite{ref:XFEL}. The finite element method calculations are needed for the
engineering design of the QWT.

The distribution of the radiation damage in the QWT is very similar to the
distribution of the energy deposition by the beam. The peak value of
displacements per atom after 5000 hours of source operation is 0.12 dpa. It was
estimated in FLUKA for room temperature. If the 0.5 dpa can be considered as an
upper limit for the damage, then the QWT can be used several years without
issues.

\section{Summary}

The positron generation and capture in the undulator-based source at 250 GeV CM
energy were simulated. Two different positron matching devices (a pulsed FC and
a QWT) downstream the target were considered. The issue of a too high peak
energy deposition in the 3.2~T FC can be reduced down to 12~J/(g pulse) by
simply increasing the aperture radius from 6.5 mm to 8 mm in case of using the
compact electron dogleg, the ILC baseline undulator $K$ value of 0.8 and the
photon collimator with an aperture radius of 3 mm upstream to the 7~mm thick
Ti6Al4V target. The required positron yield of the positron source with 1.04~T
QWT can be achieved by increasing the undulator $K$ value to 0.85. The PEDD in
the QWT is 7~J/(g~pulse), that is below the limit for iron yoke (12.5~J/g, ARMCO
Pure Steel). The peak annual radiation damage of QWT is at relatively safe level
of 0.12 dpa.  

\section{Acknowledgments}

This work was supported by the German Federal Ministry of Education and
Research, Joint Research Project R\&D Accelerator ``Positron Sources'',
Contract Number 05H15GURBA.

\end{document}